\documentclass[10pt,aps,pra,showpacs,showkeys,twocolumn,superscriptaddress]{revtex4-1}
\usepackage[dvips]{color}
\usepackage{graphicx}
\usepackage{amssymb}
\usepackage{amsmath}
\usepackage[latin1]{inputenc}
\newcommand{\be}{\begin{eqnarray}}
\newcommand{\ee}{\end{eqnarray}}
\begin{document}

\title{Entanglement dynamics via semiclassical propagators in systems of two spins}
\author{A. D. Ribeiro}
\author{R. M. Angelo}
\affiliation{Departamento de F\'{\i}sica, Universidade Federal do Paran\'a, 81531-990, Curitiba, PR, Brazil}

\begin{abstract}
We analyze the dynamical generation of entanglement in systems of two interacting spins initially prepared in a product of spin coherent states. For arbitrary time-independent Hamiltonians, we derive a semiclassical expression for the purity of the reduced density matrix as function of time. The final formula, subsidiary to the linear entropy, shows that the short-time dynamics of entanglement depends exclusively on the stability of trajectories governed by the underlying classical Hamiltonian. Also, this semiclassical measure is shown to reproduce the general properties of its quantum counterpart and give the expected result in the large spin limit. The accuracy of the semiclassical formula is further illustrated in a problem of phase exchange for two particles of spin $j$. 
\end{abstract}


\maketitle

\section{Introduction}

When two initially separated quantum systems are led to interact with each other they lose their individuality. This means that it is no longer possible to express the state of one of the systems separately from the other, i.e., they have got {\em entangled}. The relevance of these quantum correlations, which was recognized already in the early days of the Quantum Theory, nowadays dispenses with further highlights. Entanglement has definitely achieved a prominent place within the quantum phenomenology~\cite{horodecki,amicoRMP}.

In a less consensual scenario, foundational questions have been posed which try to decipher {\em if} and {\em how} entanglement manifests in the classical limit. Surprisingly, even though it is hard to conceive a classical image of entanglement at a first sight, there exists a number of works reporting on the persistence of entanglement in semiclassical regime. Although these works agree on this essential point, their approaches are clearly different in methodology, interpretation, and even on the very notion of semiclassical limit. 

In a seminal work~\cite{kyoko}, Furuya and co-authors have numerically shown that in the short-time regime entanglement behaves in accordance with the underlying classical dynamics, with accentuated differences between chaotic and regular initial conditions. A key ingredient in this approach is the use of coherent states, which are used as initial states for the dynamics as well as to furnish, through a well-defined prescription, the corresponding classical structure. The approaches of Refs.~\cite{angelo01,marcel2005} follow the same essence, though the last one focuses on systems of two spins. References~\cite{angelo04,angelo05,brumer,casati,matzkin11}, on the other hand, propose to link entanglement with entropic measures defined within classical-statistical theories. Still, some authors have investigated the semiclassical limit of entanglement (and of decoherence) by applying time-dependent perturbation theory and diagonal  approximations~\cite{prosen05,gong03,bonanca2011}.

The present work lies in the context delineated by Refs.~\cite{jacquod1,jacquod2,pra2010}. Basically, these papers employ semiclassical propagators to analyze the entanglement dynamics of bipartite quantum systems. In Jacquod's approach~\cite{jacquod1,jacquod2}, the approximation is performed using momentum and space representations simultaneously, while in our previous article~\cite{pra2010} we adopt the coherent-state representation. Although both calculations yield the same basic results, ours has the advantage of having been naturally structured to accommodate spin degrees of freedom. The aim of this contribution is to carry on this program, providing, for the first time, a semiclassical expression for entanglement dynamics of two-spin systems.

This paper is organized as follows. We start Sec.~\ref{propsc} by reviewing the main elements of the formal structure associated to the semiclassical spin-coherent-state propagator. We then introduce the time-reversal propagator and unify the formalism, this being the first contribution of this paper. With the basic ingredients at hand, we present in Sec.~\ref{semiP} our main result: a semiclassical expression for the entanglement dynamics. The formula is analyzed in Sec.~\ref{analysis} as follows. First, the canonical result~\cite{pra2010} is shown to be exactly reproduced in an appropriate limit. Second, we test the accuracy of our semiclassical result in describing the entanglement dynamics for the problem of phase coupling between two spins $j$. In Sec.~\ref{conclusion}, we present our final remarks. 

\section{Semiclassical Propagator in the spin-coherent-state representation \label{propsc}}

The development of semiclassical approximations for the quantum propagator in the coherent-state representation has a long history. It started about 30 years ago with Klauder's approach~\cite{scsp1} on the one-dimensional {\em canonical}-coherent-state propagator, $\mathrm{K}(z_{\eta},z_{\mu},T)\equiv\langle z_{\eta}| e^{-i\hat H T/\hbar}|z_{\mu}\rangle$. Subsequently, other works~\cite{scsp2,scsp3,aguiar01} substantially contributed to the understanding of the semiclassical version of $\mathrm{K}(z_{\eta},z_{\mu},T)$. In particular, Ref.~\cite{aguiar01} consists in a very detailed study of the subject and will be, therefore, the main support to our approach. Moreover, extensions of the semiclassical formula to further canonical degrees of freedom can be found in Refs.~\cite{ribeiro04,garg1}, while derivations for spin variables are given in Refs.~\cite{sscsp2,sscsp3,sscsp4,sscsp5,garg2}. Also, it is worth mentioning a result on the two-dimensional semiclassical propagator for the case where one variable is a spin and the other is canonical~\cite{ribeiro06} and a recent derivation for SU(n)-coherent-states~\cite{thiago}.

Despite this vast literature on semiclassical propagators, only recently a result has been reported~\cite{pra2010}, providing a semiclassical approximation for {\em time-reversal} propagators using the canonical states. In what follows, we extend this result by deriving a unified formula for the two-dimensional semiclassical propagator in the {\em spin-coherent-state representation}, expression which is considered as the first contribution of this paper. However, before presenting it, for the sake of completeness, we briefly review some elements of the spin-coherent-state formalism.

\subsection{Spin Coherent States}

Spin coherent states were introduced by Radcliffe~\cite{radcliffe} in direct analogy to canonical coherent states. Since then, they have become important tools in a variety of areas of physics (see Refs.~\cite{klauderb,perelomov,gilmore,gazeau} for examples and further details). 

The {\em spin coherent state} associated with a particle of spin $j$ is defined as
\begin{equation}
|s\rangle\equiv\frac{\exp{\left\{s\hat{J}_+\right\}}}{\left(1+|s|^2\right)^{j}}
|-j\rangle,
\label{spin}
\end{equation}
where the label $s$ is a complex number, $\hat{J}_+$ is the raising spin operator and $|-j \rangle$ is the lowest eigenstate of $\hat{J}_3$ with eigenvalue $-j$. The notation adopted here is such that both $s$ and $\hat{J}_+$ are dimensionless quantities. That is, in this paper the operator $\hat{\mathbf{J}}$ denotes the usual angular momentum operator divided by $\hbar$, so that its components satisfy 
\begin{equation}
[\hat{J}_1,\hat{J}_2]= i\hat{J}_3, 
\label{amcr}
\end{equation}
plus cyclic commutation relations. In terms of these states, an over-complete unity resolution can be written as
\begin{equation}
\int |s \rangle \langle s | \, d\nu(s)\equiv 1_{s},
\quad
d\nu(s) = \frac{2j+1}{\pi} \frac{ds^{(R)}ds^{(I)}}
{\left( 1 + |s |^2 \right)^{2}},
\label{NO1}
\end{equation}
where $s^{(R)}$ and $s^{(I)}$ are, respectively, the real and the imaginary parts of $s$, and the integral runs from $-\infty$ to $+\infty$. In addition, spin coherent states are, in general, non-orthogonal as can be seen in the overlap
\begin{equation}
\langle s_\eta | s_\mu \rangle = 
\frac{\left( 1 + s_\eta^* s_\mu \right)^{2j}}
{\left( 1 + |s_\eta |^2 \right)^{j}
\left( 1 + |s_\mu |^2 \right)^{j}}.
\label{NO2}
\end{equation}
It can be shown that $|s\rangle$ saturates the uncertainty relation $\langle \Delta \hat{A}^2\rangle  \langle \Delta \hat{B}^2\rangle \ge \frac{1}{4}|\langle [\hat A,\hat B]\rangle|^2+ \frac{1}{4}|\langle \{\Delta\hat A,\Delta\hat B\}\rangle|^2$~\cite{ur} for angular momentum operators, which implies that spin coherent states are minimum uncertainty states. 

\subsection{Spin Semiclassical Propagator}
 
Let the {\em forward} ($\xi=+1$) and {\em backward} ($\xi=-1$) quantum propagator in the spin-coherent-state representation be written as
\begin{equation}
\mathrm K_{\xi}\left( \mathbf s_{\eta}^*,\mathbf s_{\mu}, T \right)\equiv
\langle s_{\eta x},s_{\eta y}| e^{-i\xi\hat{H}T/\hbar} |s_{\mu x},s_{\mu y}\rangle.
\nonumber
\end{equation}
Considering the limits $j\to\infty$ and $\hbar\to0$ with the product $\hbar j$ finite, we follow Refs.~\cite{aguiar01,pra2010,ribeiro06} to obtain the semiclassical formula
\begin{equation}
K_{\xi}\left( \mathbf s_{\eta}^*,\mathbf s_{\mu}, T \right)=
\sum_{c.t.}\sqrt{\mathcal P_{\xi}}~
e^{\frac{i}{\hbar}\left(\mathcal{S_{\xi}}+\mathcal{G_{\xi}}\right)-\Lambda}.
\label{PP}
\end{equation}
The indices $x$ and $y$ in $|\mathbf s\rangle \equiv | s_x \rangle \otimes| s_y\rangle$ refer to different subsystems. We assume, for simplicity, that the spins have the same magnitude $j$, i.e., both Hilbert spaces have dimension $2j+1$. The right-hand side of Eq.~(\ref{PP}) depends only on {\em complex trajectories} governed by a Hamiltonian function $\tilde H $ (see below). In terms of auxiliary variables $\mathbf u$ and $\mathbf v$, the Hamilton equations are
\begin{equation}
\frac{\partial\tilde{H}}{\partial u_k}=
\frac{-2ij\hbar~\dot{v}_k}{\left(1+u_kv_k\right)^2} 
\quad\mathrm{and}\quad
\frac{\partial\tilde{H}}{\partial v_k}=
\frac{2ij\hbar~\dot{u}_k}{\left(1+u_kv_k\right)^2} ,
\label{emnn}
\end{equation}
where $k=x,y$ and $\tilde H(\mathbf u,\mathbf v) = \tilde H(\mathbf s, \mathbf s^*)\equiv \langle \mathbf s| \hat{H} | \mathbf s\rangle$. This equality implicitly defines the new variables through the replacement of $\mathbf s$ and $\mathbf s^*$ by $\mathbf u$ and $\mathbf v$, respectively. Trajectories contributing to Eq.~(\ref{PP}) must satisfy the boundary conditions
\begin{equation}
\begin{array}{l}
\mathbf u'= \mathbf s_\mu \quad \mathrm{and} \quad 
\mathbf v''=\mathbf s_\eta^*, \quad\mathrm{for}\quad\xi=+1,
\\
\mathbf u''= \mathbf s_\mu \quad \mathrm{and} \quad 
\mathbf v'=\mathbf s_\eta^*, \quad\mathrm{for}\quad\xi=-1.
\end{array}
\label{bbnn}
\end{equation}
In our notation, single (double) prime stands for initial (final) time. The sum in Eq.~(\ref{PP}) runs over all trajectories governed by Eqs.~(\ref{emnn}) and submitted to boundary conditions~(\ref{bbnn}).

The complex action $\mathcal{S}_\xi=\mathcal{S}_\xi(\mathbf s_\eta^*,\mathbf s_\mu,T)$ and the function $\mathcal{G}_\xi=\mathcal{G}_\xi(\mathbf s_\eta^*,\mathbf s_\mu,T)$, in Eq.~(\ref{PP}), are explicitly written as
\begin{equation}
\begin{array}{lll}
\frac{i}{\hbar}\,
\mathcal{S}_\xi&=& \displaystyle
{\xi \int_0^T\left[
j \sum_{k=x,y}
\left(\frac{u_k\dot{v}_k-v_k\dot{u}_k}{1+u_kv_k}\right)-
\frac{i}{\hbar}\,{\tilde H}\right]dt+\tilde\Lambda},\\ \\
\frac{i}{\hbar}\,
\mathcal{G}_\xi&=&
\displaystyle{-\frac{\xi }{4}\int_0^T
\sum_{k=x,y}\left[ 
\frac{\partial \dot u_k}{\partial u_k}-
\frac{\partial\dot v_k}{\partial
v_k}\right]dt}.
\end{array}
\label{S}
\end{equation}
The factors $\Lambda$ (accounting for the normalization) and $\tilde\Lambda$, appearing in Eqs.~\eqref{PP} and~\eqref{S}, respectively, are given by
\begin{equation}
\begin{array}{l}
\Lambda=\displaystyle j\sum_{k=x,y}
\ln\left[(1+|s_{\eta k}|^2)(1+|s_{\mu k}|^2) \right] ,\\ \\
\tilde\Lambda=\displaystyle j\sum_{k=x,y}
\ln\left[(1+u_k'v_k')(1+u_k''v_k'')\right].
\end{array}
\end{equation}
At last, the prefactor of Eq.~(\ref{PP}) can be written as
\begin{equation}
\mathcal P_\xi = \det \left(\frac{i}{\hbar}
\mathbf{S}^{(\xi)}_{\mathbf{s}_\mu\mathbf{s}^*_\eta}
\right)
\prod_{k=x,y}
\left(\frac{(1+u_k''v_k'')(1+u_k'v_k')}{2j}\right),
\label{pref}
\end{equation}
where
\begin{equation}
\mathbf{S}^{(\xi)}_{\mathbf{s}_\mu\mathbf{s}^*_\eta}=
\left(\begin{array}{cc}
\frac{\partial^2\mathcal S_\xi}{\partial s_{\mu x} \partial s^*_{\eta x}}&
\frac{\partial^2\mathcal S_\xi}{\partial s_{\mu x} \partial s^*_{\eta y}}\\
\frac{\partial^2\mathcal S_\xi}{\partial s_{\mu y} \partial s^*_{\eta x}}&
\frac{\partial^2\mathcal S_\xi}{\partial s_{\mu y} \partial s^*_{\eta y}}
\end{array}\right).
\end{equation}
We point out that the phase of $\mathcal P_\xi$ plays a role similar to that of the Maslov phase in the coordinate propagator. Because of the square root in Eq.~(\ref{PP}), we must track it over time and add, after each complete turn, a phase $-\pi$ to the propagator.

For future use, we differentiate $\mathcal{S}_\xi$ to get
\begin{equation}
\frac{i}{2j\hbar}
\frac{\partial \mathcal{S}_\xi}{\partial s_{\mu k}}
= \left\{\begin{array}{ll}
\displaystyle
\frac{v_k'}{1+u_k'v_k'}&,
\quad\mathrm{for}\quad\xi=+1,\\ 
\displaystyle
\frac{ v_k''}{1+u_k''v_k''}&,
\quad\mathrm{for}\quad\xi=-1,
\end{array}\right.
\label{variaS1}
\end{equation}
and
\begin{equation}
\frac{i}{2j\hbar}
\frac{\partial \mathcal{S}_\xi}{\partial s_{\eta k}^*}
= \left\{\begin{array}{ll}
\displaystyle
\frac{ u_k''}{1+u_k''v_k''}&,
\quad\mathrm{for}\quad\xi=+1,\\ 
\displaystyle
\frac{ v_k'}{1+u_k'v_k'}&,
\quad\mathrm{for}\quad\xi=-1.
\end{array}\right.
\label{variaS2}
\end{equation}
In addition, $\partial \mathcal{S}_\xi/\partial T= -\xi{\tilde H}(\mathbf u', \mathbf v')= -\xi{\tilde H}(\mathbf u'',\mathbf v'')$. As shown in Appendix~\ref{ap1}, Eqs.~(\ref{variaS1}) and (\ref{variaS2}) allow one to write second derivatives of $\mathcal{S}_\xi$ in terms of the elements of the stability matrix $\mathbf{M}$, which is defined by
\begin{equation}
\left(
\begin{array}{c}
\delta \mathbf{u}''\\\delta \mathbf{v}''
\end{array}
\right)\equiv\mathbf{M} \left(
\begin{array}{c}
\delta \mathbf{u}'\\\delta \mathbf{v}'
\end{array}
\right)=
\left(\begin{array}{cc}
\mathbf{M}_{\mathbf{uu}} & \mathbf{M}_{\mathbf{uv}}\\ 
\mathbf{M}_{\mathbf{vu}} & \mathbf{M}_{\mathbf{vv}}
\end{array}\right)
\left(
\begin{array}{c}
\delta \mathbf{u}'\\\delta \mathbf{v}'
\end{array}
\right).
\label{stabmat}
\end{equation}
It follows that in terms of $\mathbf{M}$ the prefactor reduces to
\begin{equation}
\begin{array}{l}
\mathcal P_\xi = 
\displaystyle{
\prod_{k=x,y}\left(\frac{1+u_k''v_k''}{1+u_k'v_k'}\right)}
\times\left\{
\begin{array}{ll}
\det\mathbf{M}_{\mathbf{vv}}^{-1},& \mathrm{for}~~\xi =+1\\\\
\det\mathbf{M}_{\mathbf{uu}}^{-1},& \mathrm{for}~~\xi=-1   
\end{array}
\right. ,
\end{array}
\label{prefMonod}
\end{equation}
which is clearly more appropriate for numerical purposes.

At this point, it is worth mentioning why trajectories contributing to Eq.~(\ref{PP}) are complex in general. As pointed out after Eq.~(\ref{emnn}), $\mathbf s$ and $\mathbf s^*$ were just replaced by the new variables $\mathbf u$ and $\mathbf v$, respectively. However, if one simply considers that $\mathbf u = \mathbf v^*$, a seemly natural assumption, one cannot generally find contributing trajectories to Eq.~(\ref{PP}). In fact, for both values of $\xi$, this would impose an excessive number of boundary conditions, since the evolution time $T$, and the initial ($\mathbf u',\mathbf v'$) and final ($\mathbf u'',\mathbf v''$) phase space points would be completely determined. This over-constrained problem can be circumvented by introducing the aforementioned complex trajectories, which are obtained by extending the real and imaginary parts of $\mathbf s$ to the complex plane. This procedure is equivalent to assume that $\mathbf s^*$ is no longer the complex conjugate of $\mathbf s$. Such a maneuver, whose formal support is given in Ref.~\cite{aguiar01}, justifies why $\mathbf s$ and $\mathbf s^*$ are renamed into $\mathbf u$ and~$\mathbf v$. 

Finally, it is also important to note that if, in a given instant of time $\tau$, a trajectory has only non-null real coordinates, i.e., $\mathbf u(\tau)= [\mathbf v(\tau)]^*$, then it will be always real. This can be seen as follows. If $\hat{H}$ is Hermitian, then $\langle\mathbf s|\hat H|\mathbf s\rangle = ( \langle\mathbf s|\hat H|\mathbf s\rangle)^*$, implying that $\tilde H$ can be written as a power series of the real and imaginary parts of $\mathbf {s}$, with {\em real} coefficients. Rewriting Eq.~(\ref{emnn}) in terms of $\mathbf {s}^{(R)}$ and $\mathbf {s}^{(I)}$, one may verify that real points in phase-space, namely, those for which $\mathrm{Im}\{\mathbf {s}^{(R)}\}=\mathrm{Im}\{\mathbf {s}^{(I)}\}=0$, are allowed to possess only real phase-space velocities. It follows that the motion is constrained to the real phase-space.

\section{Semiclassical entanglement in pure bipartite spin systems \label{semiP}}

The entanglement dynamics of a pure bipartite system composed of subsystems $x$ and $y$ can be quantified by the linear entropy of the reduced density matrix, 
\be
S_{\textrm{lin}}(\hat \rho_x)=1-P(\hat \rho_x),
\label{Slin}
\ee
where $\hat \rho_x=\mathrm{Tr}_y\hat \rho$, $\hat \rho=|\psi(T)\rangle\langle\psi(T)|$, and $|\psi(T)\rangle$ is the state of the system in a given instant of time $T$. The purity of the reduced density matrix $\hat \rho_x$ is given by
\begin{eqnarray}
P (\hat \rho_x)\equiv\mathrm{Tr}_x \{\hat\rho_x^2\}=\mathrm{Tr}_x 
\left\{\left[ \mathrm{Tr}_y\hat\rho(T)\right]^2\right\},
\label{purity0}
\end{eqnarray}
a positive quantity lying on the interval $[0,1]$. For pure bipartite systems $P$ is {\em symmetric}, i.e., $P(\hat{\rho}_x)=P(\hat{\rho}_y)$, and keeps equal to unity for non-interacting systems.

In what follows, we restrict our approach to situations in which the initial state $|\psi(0)\rangle$ is a product spin-coherent-state $|\mathbf s_0\rangle=|s_{0x}\rangle\otimes|s_{0y}\rangle$, so that $S_{\textrm{lin}}(\hat{\rho}_{x,y}(0))=0$. By doing so, the matrix elements of the density operator in the spin-coherent-state representation,
\begin{eqnarray}
\langle \mathbf s_\eta|\hat\rho(T)|\mathbf s_\mu\rangle 
&=&
\langle \mathbf s_\eta| e^{-i\hat HT/\hbar}|\mathbf s_0 \rangle
\langle \mathbf s_0|e^{i\hat HT/\hbar} |\mathbf s_\mu\rangle ,
\nonumber
\end{eqnarray}
for a generic time-independent Hamiltonian $\hat H$, become kernels in Eq.~(\ref{purity0}). In terms of the notation of the previous section, these elements can be semiclassically approached by 
\begin{eqnarray}
\langle \mathbf s_\eta|\hat\rho(T)
|\mathbf s_\mu\rangle_{\mathrm{semi}}
\equiv {K}_+(\mathbf s_\eta^* ,\mathbf s_0,T)~
{K}_-(\mathbf s_0^*,\mathbf s_\mu,T).
\label{rhosemi}
\end{eqnarray}
Plugging this expression into Eq.~(\ref{purity0}) and taking the traces in the spin-coherent-state representation, we readily obtain a semiclassical version of the purity,
\begin{equation}
\begin{array}{rcl}
P_{\mathrm{sc}}(T)&\equiv&
\displaystyle
\int{K}_+((w_{x}^*,s_y^*) ,\mathbf s_0,T)~
{K}_-(\mathbf s_0^*, (s_{x},s_{y}),T)
\\\\
&\times&
\displaystyle
{K}_+((s_{x}^*,w_y^*) ,\mathbf s_0,T)~
{K}_-(\mathbf s_0^*, (w_{x},w_{y}),T)\\\\
&\times&
\displaystyle
d\nu(s_y)~d\nu(w_y)~d\nu(s_x)~d\nu(w_x).
\end{array}
\label{puritysc0}
\end{equation}
As seen by Eq.~(\ref{NO1}), this integral spans the whole 8-dimensional real space composed of the real and imaginary parts of $s_x$, $s_y$, $w_x$, and $w_y$. 

Now, let us consider, for simplicity, that only one trajectory contributes to each propagator. Then, the integrand depends on four complex trajectories, each one contributing to its respective propagator and obeying distinct, though correlated, boundary conditions, namely,
\begin{equation}
\begin{array}{lll}
\mathbf u' = \mathbf s_0~\mathrm{and}~ 
\mathbf v'' =(w_{x}^*,s_y^*) ,& \mathrm{for}
&{K}_+((w_{x}^*,s_y^*) ,\mathbf s_0,T), \\
\mathbf v' = \mathbf s^*_0  ~\mathrm{and}~ 
\mathbf u'' = (s_{x},s_y),&\mathrm{for}
&{K}_-(\mathbf s_0^*,( s_{x}, s_y) ,T), \\
\mathbf u' = \mathbf s_0   ~\mathrm{and}~ 
\mathbf v'' =(s_{x}^*,w_y^*) ,& \mathrm{for}
&{K}_+((s_{x}^*,w_y^*) ,\mathbf s_0,T), \\
\mathbf v' = \mathbf s_0^* ~\mathrm{and}~ 
\mathbf  u'' =( w_{x}, w_y) ,
&\mathrm{for}
&{K}_-(\mathbf s_0^*,( w_{x}, w_y) ,T) .
\end{array}
\label{setcond0}
\end{equation}
Although integral~(\ref{puritysc0}) is rather unlikely to be analytically solved for general Hamiltonians, its structure is proper for the application of the saddle point approximation~\cite{bleistein}. As carefully discussed in Ref.~\cite{aguiar01}, it is possible to analytically extend integral~(\ref{puritysc0}) to a line integral over an 8-dimensional complex space, which is obtained by the complex extension of the real and imaginary parts of $s_x$, $s_y$, $w_x$, and $w_y$. This procedure is equivalent to working with the set ($s_x, s_x^*, s_y, s_y^*, w_x, w_x^*, w_y, w_y^*$) of eight {\em independent} complex variables. Obviously, such a change of variables implies that [see Eq.~(\ref{NO1})] 
\begin{equation}
d\nu(s) = \frac{2j+1}{\pi} \frac{ds^{(R)}ds^{(I)}}
{\left( 1 + |s |^2 \right)^{2}}= \frac{2j+1}{2 \pi i} \frac{ds\, ds^*}
{\left( 1 + s\, s^* \right)^{2}}.
\label{NO1cc}
\end{equation}

In this new scenario, the first step of the saddle point method can be directly performed. It consists in looking for critical points ($\bar s_x,\bar s_x^*,\bar s_y,\bar s_y^*,\bar w_x,\bar w_x^*,\bar w_y,\bar w_y^*$) of the integration variables. Neglecting derivatives of the terms $\mathcal{G}_{\xi}$ and $\mathcal P_\xi$, as justified in Ref.~\cite{aguiar01}, the saddle points are obtained from
\be
\begin{array}{rlll}
\displaystyle
\frac{\partial}{\partial \bar s_y^*}
&\Big[L_{\bar s_y}
+\frac{i}{\hbar} \mathcal{S}_+( (\bar w_{x}^*,\bar s_y^*),\mathbf s_0, T) 
 \Big]&=&\\
\displaystyle
\frac{\partial}{\partial \bar s_y}
&\Big[L_{\bar s_y}
+\frac{i}{\hbar}\mathcal{S}_-(\mathbf s_0^*, (\bar s_{x},\bar s_y),T) 
\Big]&=&\\
\displaystyle
\frac{\partial}{\partial \bar s_x^*}
&\Big[L_{\bar s_x}
+\frac{i}{\hbar} 
\mathcal{S}_+( (\bar s_{x}^*,\bar w_y^*),\mathbf s_0, T) 
\Big]&=&\\
\displaystyle
\frac{\partial}{\partial \bar s_x}
&\Big[L_{\bar s_x}
+\frac{i}{\hbar}\mathcal{S}_-(\mathbf s_0^*, (\bar s_{x},\bar s_y),T) 
\Big]&=&\\
\displaystyle
\frac{\partial}{\partial \bar w_y^*}
&\Big[L_{\bar w_y}
+\frac{i}{\hbar} \mathcal{S}_+( (\bar s_{x}^*,\bar w_y^*),\mathbf s_0, T) 
\Big]&=&\\
\displaystyle
\frac{\partial}{\partial \bar w_y}
&\Big[L_{\bar w_y}
+\frac{i}{\hbar}\mathcal{S}_-(\mathbf s_0^*, (\bar w_{x},\bar w_y),T) 
\Big]&=&\\
\displaystyle
\frac{\partial}{\partial \bar w_x^*}
&\Big[L_{\bar w_x}
+\frac{i}{\hbar}\mathcal{S}_+( (\bar w_{x}^*,\bar s_y^*),\mathbf s_0, T) 
 \Big]&=& \\
\displaystyle
\frac{\partial}{\partial \bar w_x}
&\Big[L_{\bar w_x}
+\frac{i}{\hbar}\mathcal{S}_-(\mathbf s_0^*, (\bar w_{x},\bar w_y),T) 
\Big]&=& 0,
\end{array}
\label{saddle_point}
\ee
where $L_{\alpha_k} = -2j\ln\left(1+\alpha_k\,\alpha_k^*\right)$, with $\alpha_k$ assuming $\bar s_x$, $\bar s_y$, $\bar w_x$ or $\bar w_y$. Using Eqs.~\eqref{variaS1} and~\eqref{variaS2} one shows that Eqs.~(\ref{saddle_point}) imply that the four critical trajectories contributing to Eq.~\eqref{puritysc0} should obey the following additional boundary conditions:
\begin{equation}
\begin{array}{lllll}
\bar u_y'' = \bar s_y &\mathrm{and}& \bar u_x'' = \bar w_x,
& \mathrm{for} &{K}_+((\bar w_{x}^*,\bar s_y^*) ,\mathbf s_0,T), \\
\bar v_y'' = \bar s^*_y  &\mathrm{and}& \bar v_x'' = \bar s^*_x,
&\mathrm{for} &{K}_-(\mathbf s_0^*,(\bar s_{x},\bar s_y) ,T), \\
\bar u_y'' = \bar w_y   &\mathrm{and}& \bar u_x'' = \bar s_x,
& \mathrm{for} &{K}_+((\bar s_{x}^*,\bar w_y^*) ,\mathbf s_0,T), \\
\bar v_y'' = \bar w^*_y  &\mathrm{and}& \bar v_x'' = \bar w^*_x,
&\mathrm{for} &{K}_-(\mathbf s_0^*,(\bar w_{x},\bar w_y) ,T) .
\end{array}
\label{setcond1}
\end{equation}

It follows from Eqs.~(\ref{setcond0}) and~(\ref{setcond1}) that the final boundary conditions of the four critical trajectories must be {\em real}, namely, $\bar{\mathbf u}'' = (\bar{\mathbf v}'')^*$. Since this implies that these trajectories have to be real for every instant of time, we conclude that the critical set is necessarily composed of four real trajectories. Because of this constraint, the initial boundary conditions of each trajectory become completely determined and, moreover, turn out to be the same. Therefore, there is no other option but to consider that all critical trajectories actually correspond to the same solution departing from $\bar{\mathbf u}'=\mathbf s_0$ and $\bar{\mathbf v}'=\mathbf s_0^*$. Clearly, this trajectory simultaneously satisfies Eqs.~(\ref{setcond0}) and~(\ref{setcond1}). 

Now, expanding Eq.~(\ref{puritysc0}) up to second order around the four critical trajectories produces
\be
P_{\mathrm{sc}} = 
\frac{\mathcal T}
{\det \bar{\mathbf{M}}_{\mathbf{u} \mathbf{u}}
\det \bar{\mathbf{M}}_{\mathbf{v} \mathbf{v}}}
\int
e^{\frac12 \delta \mathbf z^T \mathbf A\delta \mathbf z }
d\nu(\mathbf z),
\label{Ps_setsTP}
\ee
where $d\nu(\mathbf z) \equiv d\nu(s_y)~d\nu(w_y)~d\nu(s_x)~d\nu(w_x)$ and 
\be
\begin{array}{rll}
\delta \mathbf z^T &\equiv&
\left(\begin{array}{c}
\delta w_x \; \delta w_y \;
\delta s_x^* \; \delta w_y^* \;
\delta s_x \; \delta s_y \;
\delta w_x^* \; \delta s_y^* 
\end{array}\right),
\end{array}
\ee
with $\delta \beta_k = \beta_k-\bar \beta_k$. Here, $\beta$ assumes $w$ or $s$, or still their complex conjugates, while $k$ assumes $x$ or $y$. In addition,
\be
\mathcal T &\equiv&
\prod_{k=x,y}
\frac{(1+\bar u''_k \bar v''_{k})^2}
{(1+s_{0k}\,s_{0k}^*)^2}
\label{Tcal}
\ee
and the matrix $\mathbf A$ contains second derivatives of $\bar{\mathcal S}_\pm$ and $L_{\alpha_k}$. The Gaussian integral in Eq.~(\ref{Ps_setsTP}) can be exactly solved, as shown in Appendix~\ref{ap3}. Using the result~(\ref{fr}) one may rewrite Eq.~(\ref{Ps_setsTP}) as
\begin{eqnarray}
P_{\rm sc} &=& 
\frac{\mathcal T}{\sqrt{(d - d')^2 - {d''}^2}},
\label{fr1}
\end{eqnarray}
where
\begin{equation}
\begin{array}{lll}
d &=&
\det{\bar{\bf M}_{\bf uu}}\det{\bar{\bf M}_{\bf vv}}+
\det{\bar{\bf M}_{\bf uv}}\det{\bar{\bf M}_{\bf vu}},
\\
d' &=& 
\det\bar{\mathbf A}\det\bar{\mathbf B}+
\det\bar{\mathbf C}\det\bar{\mathbf D},
\\
d'' &=& 
\det\bar{\mathbf A}'\det\bar{\mathbf B}'+
\det\bar{\mathbf C}'\det\bar{\mathbf D}',
\end{array}
\nonumber
\end{equation}
with the auxiliary matrices
\begin{equation}
\begin{array}{lll}
\left(\begin{array}{cc}
\bar{\mathbf A}&\bar{\mathbf D}\\
\bar{\mathbf C}&\bar{\mathbf B}
\end{array}\right) &\equiv&
\left(\begin{array}{cccc}
1&0&0&0\\0&0&0&1\\0&0&1&0\\0&1&0&0
\end{array}\right)
\bar{\mathbf M},\\\\
\left(\begin{array}{cc}
\bar{\mathbf A}'&\bar{\mathbf D}'\\
\bar{\mathbf C}'&\bar{\mathbf B}'
\end{array}\right) &\equiv&
\left(\begin{array}{cccc}
1&0&0&0\\0&0&1&0\\0&1&0&0\\0&0&0&1
\end{array}\right)
\bar{\mathbf M}.
\end{array}
\label{ABCD}
\end{equation}
In writing $P_{\rm sc}$ in terms of these auxiliary matrices, we have used the relations
\begin{equation}
\begin{array}{lll}
\bar{\mathbf M}_{\bf uv}\bar{\mathbf M}_{\bf vv}^{-1}
&=&
\displaystyle
\frac{1}{\det\bar{\mathbf M}_{\bf vv}}
\left(\begin{array}{cc}
\det\bar{\mathbf D}&-\det\bar{\mathbf D}'\\
\det\bar{\mathbf B}'&\det\bar{\mathbf B}
\end{array}\right) ,
\\\\
\bar{\mathbf M}_{\bf vu}\bar{\mathbf M}_{\bf uu}^{-1}
&=&
\displaystyle
\frac{1}{\det\bar{\mathbf M}_{\bf uu}}
\left(\begin{array}{cc}
\det\bar{\mathbf C}&\det\bar{\mathbf A}'\\
-\det\bar{\mathbf C}'&\det\bar{\mathbf A}
\end{array}\right) ,
\end{array}
\label{MtoABCD}
\end{equation}
which can be directly verified. Equation~(\ref{fr1}) can be further simplified by noting that the determinant of matrix $\bar{\mathbf M}$ can be written as
\begin{equation}
\det\bar{\mathbf M} = d - d' - d'',
\end{equation}
so that
\begin{eqnarray}
P_{\rm sc} &=& 
\frac{\mathcal T}{\sqrt{\det\bar{\mathbf M}\left[\det\bar{\mathbf M} +2 d''\right]}}=
\left[1+\frac{2d''}{\mathcal T} \right]^{-1/2}.\,\,
\label{fr2}
\end{eqnarray}
To derive the last equation we have used the result $\det\bar{\mathbf M} =\mathcal T$, whose demonstration is left to Appendix~\ref{ap4}. 

Equation~(\ref{fr2}) is the main result of this paper. It correctly reproduces two important properties of the quantum purity for pure bipartite systems. First, through the analysis of the elements of $\bar{\mathbf M}$ one may readily verifies that $d''=0$ for  non-interacting systems. In this case, Eq.~(\ref{fr2}) reduces to $P_{\rm sc}(T)=1$ (and $S_{\textrm{lin}}(T)=0$), as expected. Second, Eq.~(\ref{fr2}) is symmetric, since it is invariant under the exchange of the indices $x$ and $y$. This can be shown by direct inspection of Eq.~\eqref{Mdet} and the elements of $d''$, 
\begin{equation}
\begin{array}{lll}
\det\bar{\mathbf A}'&=&
\displaystyle
-\frac{(1+\bar u''_x\bar v''_x)^2}{2ij\hbar}
\frac{\partial^2\bar{\mathcal S}_-}{\partial \bar{u}''_x\partial \bar{u}''_y}
\det \bar{\mathbf M}_{\bf uu} ,
\\
\det\bar{\mathbf B}'&=&
\displaystyle
-\frac{(1+\bar u''_y\bar v''_y)^2}{2ij\hbar}
\frac{\partial^2\bar{\mathcal S}_+}{\partial \bar{v}''_y\partial \bar{v}''_x}
\det \bar{\mathbf M}_{\bf vv} ,
\\
\det\bar{\mathbf C}'&=&
\displaystyle
+\frac{(1+\bar u''_y\bar v''_y)^2}{2ij\hbar}
\frac{\partial^2\bar{\mathcal S}_-}{\partial \bar{u}''_y\partial \bar{u}''_x}
\det \bar{\mathbf M}_{\bf uu} ,
\\
\det\bar{\mathbf D}'&=&
\displaystyle
+\frac{(1+\bar u''_x\bar v''_x)^2}{2ij\hbar}
\frac{\partial^2\bar{\mathcal S}_+}{\partial \bar{v}''_x\partial \bar{v}''_y}
\det \bar{\mathbf M}_{\bf vv} ,
\end{array}
\label{MltoS}
\end{equation}
which are obtained from the last of Eqs.~(\ref{SM+}) and the first of Eqs.~(\ref{SM-}), combined with Eq.~(\ref{MtoABCD}).

\section{Analysis \label{analysis}}

In this section, further arguments are given which help one to access the qualities and limitations of the semiclassical formula~\eqref{fr2} as a quantifier of entanglement dynamics.

We start by noting that Eq.~(\ref{fr2}) essentially contains correlations among elements of the stability matrix. Remarkably, this means that the onset of entanglement is exclusively determined by the stability of a trajectory departing from the center of $|\mathbf s_0\rangle$. This trajectory, which is selected by rigid boundary conditions imposed by the approximation method, is the solution of a classical structure defined by equations of motion~\eqref{emnn} and Hamiltonian $\tilde{H}=\langle\mathbf{s}|\hat{H}|\mathbf{s}\rangle$. This result is in total agreement with those reported in Refs.~\cite{jacquod1,jacquod2,pra2010} for canonical degrees of freedom and, to the best of our knowledge, is the first of this nature for systems of spins.

Also noticeable is the fact that $P_{\rm sc}$ does not depend on $\hbar$ or $j$ {\em separately}, except through $\tilde H$. A direct inspection of equations of motion \eqref{emnn}---the building blocks of $\bar{\mathbf M}$ and hence of $P_{\rm sc}$---reveals an explicit dependence only on the product $\hbar j$, which keeps finite in semiclassical regime. As a consequence, we expect our result to remain valid even in the strict classical limit, as defined by $\hbar= 0$, $j=\infty$, and $\hbar j$ finite. Moreover, one may regard this as a formal proof that entanglement must survive in the classical limit of closed pure systems.

A careful inspection of the semiclassical propagators reveals that the exclusive dependence on $\hbar j$ derives from the fact that all four contributing trajectories coalesce to a single solution. As a consequence, contributions emerging from the exponentials, which contain, separately, terms on $\hbar$ and $j$, cancel out identically as evidenced in Eq.~(\ref{Ps_setsTP}). While this simplifies the calculation, since that all functions turn out to be expanded around a single trajectory, the validity of our result gets restricted. Indeed, it seems that semiclassical approaches containing just one contributing trajectory do not contemplate more complex behaviors, as oscillations and revivals, or even longer evolution times. Usually, such features are well described in semiclassical physics only when more trajectories are considered~\cite{hellerLT, marcel, delos}. Then we expect that, in general, our derivation be valid just for short values of time, region where just one trajectory is able to reproduce quantum results. We point out that our program here was just to keep the standard steps of the saddle point method arriving at a first formula, letting improvements on the formalism to a future work. 

Finally, although the derivation of Eq.~(\ref{PP}) demands the limit $j\to\infty$, as discussed in Refs.~\cite{sscsp4,sscsp5,ribeiro06} this kind of approximation also applies for systems with spin $j=1/2$. Basically, it works because second order expansions, essence of the approximations performed, are enough to describe correctly the dynamics of spin-$\frac12$ systems. We then expect that Eq.~(\ref{fr2}) may be also applied to this class of problems.

\subsection{The canonical limit \label{Lspin}}
 
A further interesting test for our result concerns the canonical limit. According to Refs.~\cite{perelomov,gazeau}, canonical coherent states can be obtained from spin coherent states through a contraction process, which is implemented as follows. Introducing scaled quantities $s=z/\sqrt{2j}$ and $\hat J_+=\sqrt{2j}\hat a^\dagger$, one takes the limit $j\to\infty$ to get
\begin{eqnarray*}
|s\rangle
\longrightarrow
\frac{ \exp \left\{ z \hat{a}^{\dagger} \right\}} 
{\left(1+\frac{|z|^2/2}{j}\right)^j}|-j\rangle
\approx e^{ z\hat{a}^{\dagger}-\frac{1}{2}|z|^2}|0\rangle=|z\rangle,
\end{eqnarray*}
$|z\rangle$ being the well-known canonical coherent state. In addition, discarding terms smaller than $j^{-1}$, it immediately follows that
\begin{equation}
\begin{array}{rcl}
\displaystyle
j\frac{s\dot{s}^* - \dot{s}s^*}{1+s s^*}
&\longrightarrow& 
\displaystyle
\frac{1}{2} (z\dot{z}^*-\dot{z}z^*) , 
\\
\displaystyle
(1+s_\eta s_\mu^*)^j
&\longrightarrow& 
\displaystyle
\exp\left\{\frac{1}{2} z_\eta z_\mu^*\right\}, \\
\displaystyle
\frac{\partial \dot s}{\partial s} + 
\frac{\partial \dot s^*}{\partial s^*}
&\longrightarrow& 
\displaystyle
-2\frac{i}{\hbar}
\frac{\partial^2 \tilde H}{\partial z\partial z^*},
\end{array}
\label{StoCan1}
\end{equation}
and
\begin{equation}
\begin{array}{l}
\displaystyle
\det \mathbf{S}^{(\xi)}_{\mathbf{s}_\mu\mathbf{s}^*_\eta}
\prod_{k=x,y}
\frac{(1+u_k''v_k'')(1+u_k'v_k')}{2j}
\longrightarrow 
\displaystyle
\det \mathbf{S}^{(\xi)}_{\mathbf{z}_\mu\mathbf{z}^*_\eta}.
\end{array} 
\label{StoCan2}
\end{equation}
With these expressions, we convert the formalism presented in the previous section to that of the canonical case. In addition, we should be still able to recover the semiclassical purity derived in Ref.~\cite{pra2010}, which is given by
\begin{equation}
P_{\rm sc}^{\rm (can)} = \tilde{\mathcal E}^{-1/2} 
\det\bar{\mathbf M}_{\bf uu}
\det\bar{\mathbf M}_{\bf vv},
\label{canpur}
\end{equation}
where
\begin{equation}
\begin{array}{lll}
\tilde{\mathcal E}&=&
\tilde{\mathcal E}' + 
\Big[
\left(\det\bar{\mathbf M}_{\bf uu}
\det\bar{\mathbf M}_{\bf vv}-
\det\bar{\mathbf A}
\det\bar{\mathbf B}\right) \\
&\times&
\left(\det\bar{\mathbf M}_{\bf uu}
\det\bar{\mathbf M}_{\bf vv}-
\det\bar{\mathbf C}
\det\bar{\mathbf D}\right) - \tilde{\mathcal E}''
\Big]^{2},\\
\tilde{\mathcal E}'&=&
-4 \left(\det\bar{\mathbf M}_{\bf uu}
\det\bar{\mathbf M}_{\bf vv}\det\bar{\mathbf A}'
\det\bar{\mathbf B}'\right)^2,\\
\tilde{\mathcal E}''&=&
\left(
\det\bar{\mathbf A}'\right)^2
\det\bar{\mathbf B}
\det\bar{\mathbf D}-\left(
\det\bar{\mathbf A}'
\det\bar{\mathbf B}'\right)^2\\&+&
\left(
\det\bar{\mathbf B}'\right)^2
\det\bar{\mathbf A}
\det\bar{\mathbf C}.
\end{array}\nonumber
\end{equation}
In order to prove the equivalence between Eqs.~(\ref{fr1}) and~(\ref{canpur}), we use Eqs.~(\ref{MltoS}) in the limit considered to show that
\begin{equation}
\begin{array}{lll}
\displaystyle
\frac{\sqrt{-\tilde{\mathcal E}'}}
{\det\bar{\mathbf M}_{\bf uu}
\det\bar{\mathbf M}_{\bf vv}}
&=&
\displaystyle
\det\bar{\mathbf A}'\det\bar{\mathbf B}'+ 
\det\bar{\mathbf A}'\det\bar{\mathbf B}'\\
&=&
\displaystyle
\det\bar{\mathbf A}'\det\bar{\mathbf B}'+ 
\det\bar{\mathbf C}'\det\bar{\mathbf D}'\\
&=&
\displaystyle
d''
\end{array}\nonumber
\end{equation}
and
\begin{equation}
\begin{array}{lll}
\displaystyle
\frac{\sqrt{\tilde{\mathcal E}-\tilde{\mathcal E}'}}
{\det\bar{\mathbf M}_{\bf uu}
\det\bar{\mathbf M}_{\bf vv}} 
&=& 
\det\bar{\mathbf M}_{\bf uu}
\det\bar{\mathbf M}_{\bf vv}\\
&-&
\left(\det\bar{\mathbf A}\det\bar{\mathbf B}+
\det\bar{\mathbf C}\det\bar{\mathbf D} \right)
\\
&+&
\displaystyle
\frac{\left(
\det\bar{\mathbf A}'^2 - 
\det\bar{\mathbf A}\det\bar{\mathbf C}
\right)}
{\det\bar{\mathbf M}_{\bf uu}}
\\
&\times&
\displaystyle
\frac{\left(
\det\bar{\mathbf B}'^2 - 
\det\bar{\mathbf B}\det\bar{\mathbf D}
\right)}
{\det\bar{\mathbf M}_{\bf vv}}
\\
&=&
d-d',
\end{array}
\nonumber
\end{equation}
where the last equality was obtained by using the determinant of Eq.~(\ref{MtoABCD}),
\begin{equation}
\begin{array}{lll}
-\det\bar{\mathbf M}_{\bf uu}
\det\bar{\mathbf M}_{\bf vu}
&=&
\det\bar{\mathbf A}'^2 -
\det\bar{\mathbf A}
\det\bar{\mathbf C},
\\
-\det\bar{\mathbf M}_{\bf vv}
\det\bar{\mathbf M}_{\bf uv}
&=&
\det\bar{\mathbf B}'^2 - 
\det\bar{\mathbf B}
\det\bar{\mathbf D}.
\end{array}
\end{equation}
Since $\mathcal T \to 1$ in the considered limit, simple manipulations on the above expressions complete the proof of equivalence.

Another interesting byproduct of our approach emerges by taking the canonical limit in only one of the subsystems. This procedure automatically adapts our formalism---after minor modifications on Eqs.~(\ref{StoCan1}) and~(\ref{StoCan2})---to describe, for instance, spin-boson systems.

\subsection{Case study: phase coupling \label{applic}}

Let us consider two particles, $x$ and $y$, both with spins $j$, coupled to the time-independent classical magnetic field $\mathbf{B}=(0,0,B_3)$. The {\em free} Hamiltonian may be written as $\hat{H}_0=\hat{H}_0^{(x)}+\hat{H}_0^{(y)}$, where $\hat{H}_0^{(k)}=B_3\hat{J}_3^{(k)}$, for $k=x,y$. Suppose that the spins interact with each other via the coupling
\begin{eqnarray}
\hat{H} &=& \lambda\hbar\,\left[\hat{J}_3^{(x)}\otimes\hat{J}_3^{(y)}\right],
\label{HJJ3}
\end{eqnarray}
where $\lambda$ is the coupling parameter. The Heisenberg equation $i\hbar (d\hat{H}_0^{(k)}/dt)=[\hat{H}_0^{(k)},\hat{H}_0+\hat{H}]=0$ implies that there is no energy exchange between the spins. This is why Hamiltonian \eqref{HJJ3} is said to describe a {\em phase coupling}.

Since the entanglement dynamics cannot be influenced by local terms, hereafter we work only with the interaction Hamiltonian \eqref{HJJ3} instead of the total Hamiltonian $\hat{H}+\hat{H}_0$. Also, we assume that the initial state is given by $|\psi(0)\rangle=|s_{0x}\rangle \otimes|s_{0y}\rangle$, with
\begin{eqnarray}
|s_{0k}\rangle=\frac{1}{(1+|s_{0k}|^2)^j}\sum\limits_{n_k=0}^{2j}\binom{2j}{n_k}^{1/2}s_{0k}^{n_k}|-j+n_k\rangle.\quad
\end{eqnarray}
Setting $N=(1+|s_{0x}|^2)^j(1+|s_{0y}|^2)^j$ and applying conventional techniques of the quantum formalism it is straightforward to show that
\begin{equation}
\begin{array}{lll}
P(\hat{\rho}_x(T))&=&\displaystyle
\frac{1}{N^4}
\sum
\binom{2j}{n_x}\binom{2j}{n_x'}\binom{2j}{n_y}\binom{2j}{n_y'} 
\\ &\times&
|s_{0x}|^{2\sigma_x}|s_{0y}|^{2\sigma_y}
e^{-i\lambda\, T\,\delta_x\, \delta_y},
\end{array}
\label{Px}
\end{equation}
where $\delta_k\equiv n_k-n_k'$, $\sigma_k\equiv n_k+n_k'$, and the sum is over $n_x, n_y, n_x', n_y'$, running from 0 to $\infty$. This result equals $P(\hat{\rho}_y(T))$ since it is clearly invariant by the exchange of the indices $x$ and $y$. 

In order to establish contact with the semiclassical result, we compute the short-time expression for the entanglement generation. By expanding the result \eqref{Px} up to second order in time we obtain
\begin{eqnarray}
S_{\text{lin}}(T)\cong \left[\frac{\sqrt{8}\,|s_{0x}|\,|s_{0y}|\,j\lambda\, T}{(1+|s_{0x}|^2)(1+|s_{0y}|^2)}\right]^2.
\label{applF}
\end{eqnarray}
As anticipated by the discussion of Sec.~\ref{Lspin}, we expect this result to reproduce the canonical one under the parametrization $s_{0k}=z_{0k}/\sqrt{2j}$ followed by the limit $j\to \infty$. Evaluating the above expression in these terms, we obtain that
\begin{eqnarray}
\lim\limits_{j\to \infty}S_{\text{lin}}(T)\cong 2\,|z_{0x}|^2\,|z_{0y}|^2\,(\lambda T)^2,
\end{eqnarray}
which indeed yields a result equivalent in structure to that obtained in Ref.~\cite{pra2010} for a system of two oscillators. 

To apply the semiclassical formalism to this system, we first find the classical Hamiltonian associated to Eq.~(\ref{HJJ3}):
\begin{equation}
\begin{array}{lll}
\tilde{H}(\mathbf u, \mathbf v) &=& 
\displaystyle
\langle \mathbf v|
\lambda\hbar\left[
\hat{J}_3^{(x)}\otimes\hat{J}_3^{(y)}
\right]
|\mathbf u\rangle \\\\
&=&
\displaystyle
\lambda \hbar j^2
\left(\frac{1-u_xv_x}{1+u_xv_x}\right)\left(\frac{1-u_yv_y}{1+u_yv_y}\right).
\end{array}
\end{equation}
Equations of motion~(\ref{emnn}) result in
\begin{equation}
\left(\begin{array}{c}
\dot u_x \\\dot u_y \\\dot v_x \\\dot v_y 
\end{array}\right)
=
\left(\begin{array}{cccc}
\lambda_x &0&0&0 \\
0&\lambda_y&0&0 \\
0&0&-\lambda_x&0\\
0&0&0&-\lambda_y
\end{array}\right)
\left(\begin{array}{c}
u_x \\u_y \\v_x \\v_y 
\end{array}\right),
\end{equation}
where $\lambda_x = i\lambda j \left( \frac{1-u_yv_y}{1+u_yv_y}\right)$ and $\lambda_y =i\lambda j \left( \frac{1-u_xv_x}{1+u_xv_x}\right)$. It is clear that both $u_xv_x$ and $u_yv_y$ are constants of motion. Then trajectories are readily obtained in terms of their initial conditions,
\begin{equation}
\begin{array}{ll}
u_x (t) = u_x' e^{\lambda_x t},
&u_y (t) = u_y' e^{\lambda_y t},\\
v_x (t) = v_x' e^{-\lambda_x t},
&v_y (t) = v_y' e^{-\lambda_y t}.
\end{array}
\end{equation}
From them, and remembering that $\lambda_x=\lambda_x(u_y',v_y')$ and $\lambda_y=\lambda_y(u_x',v_x')$, the stability matrix is straightforwardly written as $\mathbf{M}=\mathbf{M}_1\mathbf{M}_2$, where
\begin{equation}
\mathbf{M}_1 =2t
\left(\begin{array}{cccc}
\lambda_x e^{\lambda_x t }&0&0&0 \\
0&\lambda_ye^{\lambda_y t }&0&0 \\
0&0&\lambda_xe^{-\lambda_x t}&0\\
0&0&0&\lambda_ye^{-\lambda_y t}
\end{array}\right)
\nonumber
\end{equation}
and
\begin{equation}
\mathbf{M}_2 =
\left(\begin{array}{cccc}
\frac{1}{2\lambda_x t}&\frac{-u_x'v_y'}{1-{u_y'}^2{v_y'}^2}&
0&\frac{-u_x'u_y'}{1-{u_y'}^2{v_y'}^2} \\
\frac{-u_y'v_x'}{1-{u_x'}^2{v_x'}^2}&\frac{1}{2\lambda_y t}&
\frac{-u_y'u_x'}{1-{u_x'}^2{v_x'}^2}&0 \\
0&\frac{v_x'v_y'}{1-{u_y'}^2{v_y'}^2}&
\frac{1}{2 \lambda_x t}&\frac{v_x'u_y'}{1-{u_y'}^2{v_y'}^2}\\
\frac{v_y'v_x'}{1-{u_x'}^2{v_x'}^2}&0&
\frac{v_y'u_x'}{1-{u_x'}^2{v_x'}^2}&\frac{1}{2 \lambda_y t}
\end{array}\right).
\nonumber
\end{equation}
Then, as for this system $\mathcal T$ amounts to 1, and
\begin{equation}
\begin{array}{ll}
\det\mathbf A' = \frac{2v_x'v_y' \lambda_x t}{1-{u_y'}^2{v_y'}^2}, &\quad
\det\mathbf B' = \frac{-2u_x'u_y' \lambda_y t}{1-{u_x'}^2{v_x'}^2},\\
\det\mathbf C' = \frac{-2v_y'v_x' \lambda_y t}{1-{u_x'}^2{v_x'}^2}, &\quad
\det\mathbf D' = \frac{2u_x'u_y' \lambda_x t}{1-{u_y'}^2{v_y'}^2},
\end{array}
\end{equation}
we finally find that
\begin{equation}
\begin{array}{lll}
P_{\rm sc}(T) &=& [1+2d'']^{-1/2} \\
&=& \displaystyle
\left[1-\frac{16u'_xv'_xu'_yv'_y\lambda_x \lambda_y T^2}
{(1-{u_x'}^2{v_x'}^2)(1-{u_y'}^2{v_y'}^2)}\right]^{-1/2}\\
&\approx& \displaystyle
1- \left[\frac{\sqrt8\,|s_{0x}|\,|s_{0y}|\,j \,\lambda \,T}
{(1+|s_{0x}|^2)(1+|s_{0y}|^2)}\right]^2,
\end{array}
\end{equation}
which agrees with the quantum result~(\ref{applF}).

This case study highlights the major difficulty of our approach: the semiclassical formula applies accurately only in the short-time regime. Nevertheless, this is not really surprising. As pointed out above, it is well-known that quantum phenomena can be well described semiclassically only via many contributing trajectories. As we have seen, this is not the case here. Actually, this turns out to be one of the next challenging question in the context drawn so far: How to improve the semiclassical formula so as to correctly describe the entanglement dynamics for longer times?

\section{Final Remarks \label{conclusion}}

In summary, this paper is concerned with autonomous systems of two spins $j$ prepared in a product of spin-coherent-states. We looked at the entanglement dynamics as quantified by the linear entropy---or its kernel, the quantum purity---as a function of time. A semiclassical approximation for the purity was derived by replacing exact propagators by their semiclassical versions. The calculation, which employed the saddle point method to analytically solve the integrals, produced the semiclassical expression~(\ref{fr2}), the main result of this paper. This formula allows one to express the onset of entanglement in terms of a classical structure, defined by a Hamiltonian function, equations of motions, and a set of boundary conditions involving the initial conditions. The semiclassical time-reversal spin-coherent-state propagator~(\ref{PP}) is another original derivation of this work.

The adequacy of our results was illustrated by some important analytical tests. First, the semiclassical purity was shown to be symmetric. This property, which is not trivially reproduced by classical entropic measures~\cite{angelo04,casati}, indicates that our formula does capture the quantum essence of entanglement. Interestingly, however, the resulting structure is shown not to importantly depend on $\hbar$ or $j$ separately. This constitutes a symptom of the fact that the semiclassical result should be accurate only in the short-time regime. Second, it was shown that the semiclassical purity correctly recovers the canonical result~\cite{pra2010} in the large-spin limit. We concluded the tests with a case study which confirmed the accuracy of our semiclassical result in the regime of short times.

Finally, it is worth noting that our results and conclusions are in consonance with many others reported for canonical degrees of freedom~\cite{kyoko,angelo01,jacquod1,pra2010}, especially in what regards the link between entanglement dynamics and stability of underlying classical structures. A natural continuation of this paper includes the improvement of the semiclassical formula so as to reproduce the exact entanglement dynamics in regimes of longer values of time. Work on this topic is now in progress.

\acknowledgments
A.D.R. and R.M.A. acknowledge financial support from INCT-IQ (CNPq/Brazil).

\appendix
\section{Elements of the stability matrix and second derivatives of the action \label{ap1}} 

In this appendix we derive relations between elements of the stability matrix $\mathbf M$, defined by Eq.~(\ref{stabmat}), and second derivatives of the complex action $\mathcal{S}_\xi$, defined by Eq.~(\ref{S}). We start by performing variations on both sides of Eqs.~(\ref{variaS1}) and~(\ref{variaS2}). Dealing first with $\xi=+1$, we get
\begin{equation}
\left[ \mbox{\boldmath{$\Gamma$}}_{+} +
\left(\begin{array}{cc}
\mathbf{A}_+&0\\ 0&\mathbf{C}_+
\end{array}\right)
\right]
\left(
\begin{array}{c}
\delta \mathbf{u}'\\\delta \mathbf{v}''
\end{array}
\right)=
\left(\begin{array}{cc}
0&\mathbf{B}_+\\ \mathbf{D}_+&0
\end{array}\right)
\left(
\begin{array}{c}
\delta \mathbf{u}''\\\delta \mathbf{v}'
\end{array}
\right),
\label{d2+}
\end{equation}
where 
\begin{eqnarray}
\mbox{\boldmath{$\Gamma$}}_{+} &\equiv&
\left(\begin{array}{cc}
\mathbf{S}^{(+)}_{\mathbf{u'u'}} & \mathbf{S}^{(+)}_{\mathbf{u'v''}}\\ 
\mathbf{S}^{(+)}_{\mathbf{v''u'}} & \mathbf{S}^{(+)}_{\mathbf{v''v''}}
\end{array}\right),\nonumber \\
\mathbf{S}^{(+)}_{\mathbf{ab}} &\equiv&
\left(\begin{array}{cc}
\frac{\partial^2\mathcal S_+}{\partial a_x\partial b_x} & 
\frac{\partial^2\mathcal S_+}{\partial a_x\partial b_y} \\
\frac{\partial^2\mathcal S_+}{\partial a_y\partial b_x} & 
\frac{\partial^2\mathcal S_+}{\partial a_y\partial b_y} 
\end{array}\right),\nonumber
\end{eqnarray}
with $\mathbf{a}$ and $\mathbf{b}$ assuming  $\mathbf{u}'$ or $\mathbf{v}''$, and
\begin{eqnarray*}
\mathbf A_+ &\equiv& -2ij\hbar
\left(\begin{array}{cc}
\frac{{v'_x}^2}{(1+u'_xv'_x)^2}&0 \\ 
0& \frac{{v'_y}^2}{(1+u'_yv'_y)^2}
\end{array}\right),
\\
\mathbf B_+ &\equiv& -2ij\hbar
\left(\begin{array}{cc}
\frac{1}{(1+u'_xv'_x)^2}&0 \\ 
0& \frac{1}{(1+u'_yv'_y)^2}
\end{array}\right),
\\
\mathbf C_+ &\equiv& -2ij\hbar
\left(\begin{array}{cc}
\frac{{u''_x}^2}{(1+u''_xv''_x)^2}&0 \\ 
0& \frac{{u''_y}^2}{(1+u''_yv''_y)^2}
\end{array}\right),
\\
\mathbf D_+ &\equiv& -2ij\hbar
\left(\begin{array}{cc}
\frac{1}{(1+u''_xv''_x)^2}&0 \\ 
0& \frac{1}{(1+u''_yv''_y)^2}
\end{array}\right). \label{defmat+}
\end{eqnarray*}
Rearranging Eq.~(\ref{d2+}), so as to write the final displacements $\delta \mathbf{u}''$ and $\delta \mathbf{v}''$ as a function of the initial ones $\delta \mathbf{u}'$ and $\delta \mathbf{v}'$, and comparing it with Eq.~(\ref{stabmat}) lead to
\begin{equation}
\begin{array}{lll}
\mathbf{M_{uu}}&=&
\mathbf{D}_+^{-1}\left\{\mathbf{S}^{(+)}_{\mathbf{v''u'}}
-\tilde{\mathbf{C}}_+
\left[\mathbf{S}^{(+)}_{\mathbf{u'v''}}\right]^{-1}
\tilde{\mathbf{A}}_+\right\},\\
\mathbf{M_{uv}}&=&
\mathbf{D}_+^{-1}\tilde{\mathbf{C}}_+
\left[\mathbf{S}^{(+)}_{\mathbf{u'v''}}\right]^{-1}
\mathbf{B_+},\\
\mathbf{M_{vu}}&=&
-\left[\mathbf{S}^{(+)}_{\mathbf{u'v''}}\right]^{-1}
\tilde{\mathbf{A}}_+,\\
\mathbf{M_{vv}}&=&
\left[\mathbf{S}^{(+)}_{\mathbf{u'v''}}\right]^{-1}\mathbf{B}_+ ,
\end{array}\label{MS+}
\end{equation}
where $\tilde{\mathbf{A}}_+ \equiv \mathbf{S}^{(+)}_{\mathbf{u'u'}}+\mathbf{A}_+$ and $\tilde{\mathbf{C}}_+ \equiv \mathbf{S}^{(+)}_{\mathbf{v''v''}}+\mathbf{C}_+$. Inverting these relations, one shows that
\begin{equation}
\begin{array}{lll}
\mathbf{S}^{(+)}_{\mathbf{u'u'}} &=& 
-\mathbf{B}_+\mathbf{M}_{\mathbf{vv}}^{-1}\mathbf{M}_{\mathbf{vu}}-\mathbf{A}_+, \\
\mathbf{S}^{(+)}_{\mathbf{u'v''}}&=& 
\mathbf{B}_+\mathbf{M}_{\mathbf{vv}}^{-1} ,\\
\mathbf{S}^{(+)}_{\mathbf{v''u'}}&=& 
\mathbf{D}_+\big[\mathbf{M}_{\mathbf{uu}}-
\mathbf{M}_{\mathbf{uv}}\mathbf{M}_{\mathbf{vv}}^{-1}
\mathbf{M}_{\mathbf{vu}}
\big],\\
\mathbf{S}^{(+)}_{\mathbf{v''v''}}&=& 
\mathbf{D}_+\mathbf{M}_{\mathbf{uv}}\mathbf{M}_{\mathbf{vv}}^{-1}-\mathbf{C}_+.
\end{array} \label{SM+}
\end{equation}

Analogous relations can be found for $\xi=-1$. Differentiating Eqs.~(\ref{variaS1}) and~(\ref{variaS2}), we find that
\begin{equation}
\left[ \mbox{\boldmath{$\Gamma$}}_{-}+
\left(\begin{array}{cc}
\mathbf{A}_-&0\\ 0&\mathbf{C}_-
\end{array}\right)
\right]
\left(
\begin{array}{c}
\delta \mathbf{u}''\\\delta \mathbf{v}'
\end{array}
\right)=
\left(\begin{array}{cc}
0&\mathbf{B}_-\\ \mathbf{D}_-&0
\end{array}\right)
\left(
\begin{array}{c}
\delta \mathbf{u}'\\\delta \mathbf{v}''
\end{array}
\right),
\label{d2-}
\end{equation}
where 
\begin{eqnarray}
\mbox{\boldmath{$\Gamma$}}_{-} &\equiv&
\left(\begin{array}{cc}
\mathbf{S}^{(-)}_{\mathbf{u''u''}} & \mathbf{S}^{(-)}_{\mathbf{u''v'}}\\ 
\mathbf{S}^{(-)}_{\mathbf{v'u''}} & \mathbf{S}^{(-)}_{\mathbf{v'v'}}
\end{array}\right), \nonumber \\
\mathbf{S}^{(-)}_{\mathbf{ab}} &\equiv&
\left(
\begin{array}{cc}
\frac{\partial^2\mathcal S_-}{\partial a_x\partial b_x} & 
\frac{\partial^2\mathcal S_-}{\partial a_x\partial b_y} \\
\frac{\partial^2\mathcal S_-}{\partial a_y\partial b_x} & 
\frac{\partial^2\mathcal S_-}{\partial a_y\partial b_y} 
\end{array}
\right),\nonumber
\end{eqnarray}
with $\mathbf a$ and $\mathbf b$ now assuming $\mathbf u''$ or $\mathbf v'$, and
\begin{eqnarray*}
\mathbf{A}_- &\equiv& -2ij\hbar
\left(\begin{array}{cc}
\frac{{v''_x}^2}{(1+u''_xv''_x)^2}&0\\
0&\frac{{v''_y}^2}{(1+u''_yv''_y)^2}
\end{array}\right),\\
\mathbf{B}_- &\equiv& -2ij\hbar
\left(\begin{array}{cc}
\frac{1}{(1+u''_xv''_x)^2}&0\\
0&\frac{1}{(1+u''_yv''_y)^2}
\end{array}\right),\\
\mathbf{C}_- &\equiv& -2ij\hbar
\left(\begin{array}{cc}
\frac{{u'_x}^2}{(1+u'_xv'_x)^2}&0\\
0&\frac{{u'_y}^2}{(1+u'_yv'_y)^2}
\end{array}\right),\\
\mathbf{D}_- &\equiv& -2ij\hbar
\left(\begin{array}{cc}
\frac{1}{(1+u'_xv'_x)^2}&0\\
0&\frac{1}{(1+u'_yv'_y)^2}
\end{array}\right).
\label{defmat-}
\end{eqnarray*}
Manipulating Eq.~(\ref{d2-}) in a convenient way, we get
\begin{equation}
\begin{array}{lll}
\mathbf{M_{uu}} &=&\left[\mathbf{S}^{(-)}_{\mathbf{v'u''}}\right]^{-1}\mathbf{D}_-,\\
\mathbf{M_{uv}} &=&-\left[\mathbf{S}^{(-)}_{\mathbf{v'u''}}\right]^{-1} \tilde{\mathbf{C}}_- ,\\
\mathbf{M_{vu}} &=&\mathbf{B}_-^{-1}\tilde{\mathbf{A}}_-
\left[\mathbf{S}^{(-)}_{\mathbf{v'u''}}\right]^{-1}
\mathbf{D}_-,\\
\mathbf{M_{vv}} &=& \mathbf{B}_-^{-1}\left\{\mathbf{S}^{(-)}_{\mathbf{u''v'}} 
-\tilde{\mathbf{A}}_- 
\left[\mathbf{S}^{(-)}_{\mathbf{v'u''}}\right]^{-1}
\tilde{\mathbf{C}}_-\right\},
\end{array}\label{MS-}
\end{equation}
where $\tilde{\mathbf{A}}_-\equiv\mathbf{S}^{(-)}_{\mathbf{u''u''}}+\mathbf{A}_-$ and $\tilde{\mathbf{C}}_- \equiv \mathbf{S}^{(-)}_{\mathbf{v'v'}}+\mathbf{C}_-$. Inverting them leads to 
\begin{equation}
\begin{array}{lll}
\mathbf{S}^{(-)}_{\mathbf{u''u''}} &=& 
\mathbf{B}_-\mathbf{M}_{\mathbf{vu}}\mathbf{M}_{\mathbf{uu}}^{-1}-\mathbf{A}_-,\\
\mathbf{S}^{(-)}_{\mathbf{u''v'}} &=&\mathbf{B}_-\left[\mathbf{M}_{\mathbf{vv}}-
\mathbf{M}_{\mathbf{vu}}\mathbf{M}_{\mathbf{uu}}^{-1}
\mathbf{M}_{\mathbf{uv}}
\right] ,\\
\mathbf{S}^{(-)}_{\mathbf{v'u''}} &=& \mathbf{D}_-\mathbf{M}_{\mathbf{uu}}^{-1} ,\\
\mathbf{S}^{(-)}_{\mathbf{v'v'}} &=&-\mathbf{D}_-\mathbf{M}_{\mathbf{uu}}^{-1}\mathbf{M}_{\mathbf{uv}}-\mathbf{C}_-  .
\end{array} \label{SM-}
\end{equation}

Equations (\ref{MS+}), (\ref{SM+}), (\ref{MS-}), and (\ref{SM-}) establish the intended connection between elements of the stability matrix and second derivatives of the action. In particular, they prove the equivalence between Eqs.~(\ref{pref}) and~(\ref{prefMonod}), provided that we identify $\det \mathbf S_{\mathbf s_\mu \mathbf s_\eta^*}^{(+)}$ and $\det \mathbf S_{\mathbf s_\mu \mathbf s_\eta^*}^{(-)}$ with  $\det \mathbf S_{\mathbf u' \mathbf v''}^{(+)}$ and $\det \mathbf S_{\mathbf v' \mathbf u''}^{(-)}=\det \mathbf S_{\mathbf u'' \mathbf v'}^{(-)}$, respectively.

\section{Gaussian integral \label{ap3}}

In this appendix we solve the Gaussian integral
\begin{eqnarray*}
\mathcal{I}\equiv\int
e^{\frac12 \, \delta \mathbf z^T \mathbf A\,\delta \mathbf z}
\, d\nu(\mathbf{z}),
\end{eqnarray*}
which the semiclassical purity $P_{\rm sc}$ depends on, as shown in Eq.~(\ref{Ps_setsTP}). While $\delta \mathbf z$ and $d\nu(\mathbf{z})$ are defined in the main text, the $8\times8$ matrix $\mathbf A$ is composed of the following $4\times4$ blocks
\begin{equation}
\begin{array}{l}
\mathbf A_{11} \equiv
\left(\begin{array}{cccc}
\mathbf C_{\bar w^*} + \frac{i}{\hbar}\bar{\mathbf{S}}_{\mathbf u''\mathbf u''}^{(-)} & 
\mathbf B_{\bar w_y}\\
\mathbf B_{\bar w_y} & 
\mathbf C_{\bar s} + \frac{i}{\hbar}\bar{\mathbf{S}}_{\mathbf v''\mathbf v''}^{(+)} 
\end{array}\right),\\\\
\mathbf A_{22} \equiv
\left(
\begin{array}{cccc}
\mathbf C_{\bar s^*} + \frac{i}{\hbar}\bar{\mathbf{S}}_{\mathbf u''\mathbf u''}^{(-)} &
\mathbf B_{\bar s_y}\\
\mathbf B_{\bar s_y}&
\mathbf C_{\bar w} + \frac{i}{\hbar}\bar{\mathbf{S}}_{\mathbf v''\mathbf v''}^{(+)} 
\end{array} \right),\\\\
\mathbf A_{12} \equiv
\left(
\begin{array}{cccc}
0 & \mathbf B_{\bar w_x} \\
\mathbf B_{\bar s_x}  & 0
\end{array} \right),~\mathrm{and}~
\mathbf A_{21} \equiv
\left(
\begin{array}{cc}
0&\mathbf B_{\bar s_x}\\
\mathbf B_{\bar w_x}&0
\end{array} \right),
\end{array}
\nonumber
\end{equation}
where $\mathbf C_{\bar\alpha} = -\bar\alpha_x^2\mathbf B_{\bar\alpha_x}-\bar\alpha_y^2\mathbf B_{\bar\alpha_y} $, 
\begin{equation}
\mathbf B_{\bar\alpha_x}\equiv
\left(\begin{array}{cc}
\frac{-2j}{(1+\bar\alpha_x\,\bar\alpha_x^*)^2}&0\\0&0
\end{array} \right)~\mathrm{and}~
\mathbf B_{\bar\alpha_y}\equiv
\left(\begin{array}{cc}
0&0\\0&\frac{-2j}{(1+\bar\alpha_y\,\bar\alpha_y^*)^2}
\end{array} \right),
\nonumber
\end{equation}
with $\alpha$ assuming $s$, $s^*$, $w$, and $w^*$. In Appendix~\ref{ap1}, second derivatives of the actions $\mathcal S_\pm$ are written in terms of the stability matrix ${\mathbf M}$ of the pertinent trajectory. Using Eqs.~(\ref{SM+}) and~(\ref{SM-}), and recalling that the trajectory associated to $\bar{\mathcal S}_+$ is identical to that associated to $\bar{\mathcal S}_-$, we rewrite the above matrices as
\begin{eqnarray*}
\mathbf{A}_{11} =\mathbf{A}_{22} &=&
\left(\begin{array}{cc}
- \mathbf S\,\bar{\bf M}_{\bf vu}\bar{\bf M}_{\bf uu}^{-1} & \mathbf S\,\mathbf{R}_y\\
\mathbf S\,\mathbf{R}_y& - \mathbf S\,\bar{\bf M}_{\bf uv}\bar{\bf M}_{\bf vv}^{-1}
\end{array} \right),\\
\mathbf{A}_{12} =\mathbf{A}_{21} &=&
\left(\begin{array}{cc}
0&\mathbf S\,\mathbf{R}_x\\
\mathbf S\,\mathbf{R}_x&0
\end{array} \right),
\end{eqnarray*}
where
\begin{eqnarray*}
&&\mathbf{R}_x=
\left(\begin{array}{cc}
1&0\\
0&0
\end{array} \right),\quad
\mathbf{R}_y=
\left(\begin{array}{cc}
0&0\\
0&1
\end{array} \right),\;\;\mathrm{and}\\
&&
\mathbf S =
\left(\begin{array}{cc}
\frac{-2j}{(1+\bar u_x''\bar v_x'')^2}&0\\
0&\frac{-2j}{(1+\bar u_y''\bar v_y'')^2}
\end{array} \right).
\end{eqnarray*}
With these arrangements, the determinant of ${\bf A}$ can be straightforwardly calculated, resulting that
\begin{eqnarray*}
\det \mathbf A = \left( \det \mathbf S \right)^4\left[ a_1^2 - a_2^2\right],
\end{eqnarray*}
where 
\begin{eqnarray*}
a_1 &=& 1+ 
\det\bar{\bf M}_{\bf vu}\det\bar{\bf M}^{-1}_{\bf uu}
\det\bar{\bf M}_{\bf uv}\det\bar{\bf M}^{-1}_{\bf vv}
\\&-&
\mathbf h_x^T\, \bar{\bf M}_{\bf vu}\,
\bar{\bf M}_{\bf uu}^{-1}\,\mathbf h_x\, 
\mathbf h_x^T \,\bar{\bf M}_{\bf uv}\,
\bar{\bf M}_{\bf vv}^{-1}\,\mathbf h_x
\\&-&
\mathbf h_y^T\,\bar{\bf M}_{\bf vu}\,
\bar{\bf M}_{\bf uu}^{-1}\,\mathbf h_y\,
\mathbf h_y^T\,\bar{\bf M}_{\bf uv}\,
\bar{\bf M}_{\bf vv}^{-1}\,\mathbf h_y\,,
\\
a_2 &=& 
\mathbf h_x^T\,\bar{\bf M}_{\bf vu}\,
\bar{\bf M}_{\bf uu}^{-1}\,\mathbf h_y \,
\mathbf h_y^T\,\bar{\bf M}_{\bf uv}\,
\bar{\bf M}_{\bf vv}^{-1}\,\mathbf h_x
\\&+&
\mathbf h_y^T\,\bar{\bf M}_{\bf vu}\,
\bar{\bf M}_{\bf uu}^{-1}\,\mathbf h_x\,
\mathbf h_x^T\,\bar{\bf M}_{\bf uv}\,
\bar{\bf M}_{\bf vv}^{-1}\,\mathbf h_y,
\end{eqnarray*}
with $\mathbf{h}_x^T\equiv(1,0)$ and $\mathbf{h}_y^T\equiv(0,1)$. Using Eq.~(\ref{NO1cc}) and $\frac{2j+1}{2j}\approx 1$, which becomes exact in the limit considered, we finally find that
\begin{equation}
\begin{array}{lll}
\mathcal{I}&=&
\displaystyle
\frac{(2j+1)^4\left( \det \mathbf A\right)^{-1/2}}
{(1+\bar u_x''\bar v_x'')^4(1+\bar u_y''\bar v_y'')^4} 
\approx \sqrt{\frac{1}{a_1^2 - a_2^2}}.
\end{array}
\label{fr}
\end{equation}

\section{Determinant of the stability matrix \label{ap4}}

Here we derive an expression for the determinant of ${\mathbf{M}}$ [Eq.~(\ref{stabmat})], the stability matrix associated to the classical trajectory involved in the calculation of $P_{\rm sc}$. Because of the symplectic structure of {\em canonical} Hamilton's Equations, the determinant of the stability matrix is constant and equals to 1 (see, for instance, Ref.~\cite{CM}). However, for the spin equations of motion~(\ref{emnn}), the above no longer holds. Our strategy to compute $\det{\mathbf{M}}$ consists in introducing a new set of canonical variables $q_x,\,p_x,\,q_y$ and $p_y$~\cite{aguiar92,ribeiroJMP07}, for which $\det{\mathbf{M}}_{\rm can}=1$. Then, from the relation between the two set of variables, $\det{\mathbf{M}}$ can be determined.

Assuming that $u_k=u_k(q_k,p_k)$ and $v_k=v_k(q_k,p_k)$, for $k=x,y$, implies that
\begin{equation}
\delta \mathbf{w} = \mathbf{T}~ \delta \mathbf{r},\label{zzss}
\end{equation}
where we have defined $\delta \mathbf{w}^T\equiv \left(\delta u_x \; \delta u_y \; \delta v_x \; \delta v_y \right)$ and $\delta \mathbf{r}^T\equiv \left(\delta q_x \; \delta q_y \; \delta p_x \; \delta p_y \right)$. Non-null elements of $\mathbf T$ are given by the relations
\begin{equation}
\begin{array}{ll}
t_{11} \equiv \frac{\partial u_x}{\partial q_x}
= J_x \frac{\partial p_x}{\partial v_x},  
& 
t_{13} \equiv \frac{\partial u_x}{\partial p_x}
= -J_x \frac{\partial q_x}{\partial v_x}  , 
\\
t_{22} \equiv \frac{\partial u_y}{\partial q_y}
= J_y \frac{\partial p_y}{\partial v_y}  ,  
& 
t_{24} \equiv \frac{\partial u_y}{\partial p_y}
= -J_y \frac{\partial q_y}{\partial v_y}  , 
\\
t_{31} \equiv \frac{\partial v_x}{\partial q_x}
= -J_x \frac{\partial p_x}{\partial u_x}  , 
& 
t_{33} \equiv \frac{\partial v_x}{\partial p_x}
= J_x  \frac{\partial q_x}{\partial u_x} , 
\\
t_{42} \equiv \frac{\partial v_y}{\partial q_y}
= -J_y \frac{\partial p_y}{\partial u_y}  , 
& 
t_{44} \equiv \frac{\partial v_y}{\partial p_y}
= J_y  \frac{\partial q_y}{\partial u_y} ,
\end{array}
\label{eqc1}
\end{equation}
where $J_x\equiv t_{11} t_{33} -t_{13} t_{31} $, $J_y\equiv t_{22} t_{44} -t_{24} t_{42}$, and the last term of each equation is obtained by inverting Eq.~(\ref{zzss}).

By demanding $q_k$ and $p_k$ to be canonical coordinates, one must require that
\begin{equation}
\begin{array}{lll}
\dot q_k &=& \displaystyle
\frac{\partial q_k}{\partial u_k}\dot u_k + \frac{\partial q_k}
{\partial v_k}\dot v_k
= \frac{(1+u_kv_k)^2}{2ij\hbar} \{ q_k,\tilde H\}_{u_k,v_k}\\
&=&  \displaystyle\frac{(1+u_kv_k)^2}{2ij\hbar} \{ q_k,p_k\}_{u_k,v_k}  
\frac{\partial\tilde{\mathrm H}}{\partial p_k}
=\frac{\partial \tilde{\mathrm H}}{\partial p_k},
\\ \\
\dot p_k &=& \displaystyle
\frac{\partial p_k}{\partial u_k}\dot u_k + 
\frac{\partial p_k}{\partial v_k}\dot v_k
= \frac{(1+u_kv_k)^2}{2ij\hbar}  \{ p_k,\tilde H\}_{u_k,v_k}\\
&=& \displaystyle \frac{(1+u_kv_k)^2}{2ij\hbar} \{ q_k,p_k\}_{v_k,u_k} \frac{\partial \tilde{\mathrm H}}{\partial q_k}
=-\frac{\partial \tilde{\mathrm H}}{\partial q_k},
\end{array} \label{ec}
\end{equation}
where Eq.~(\ref{emnn}) was used to eliminate the time derivative. In these relations, $\tilde{\mathrm H}(q_x,q_y,p_x,p_y) $ amounts to
\begin{equation}
\tilde{H}[u_x(q_x,p_x),u_y(q_y,p_y),v_x(q_x,p_x),v_y(q_y,p_y)]. 
\nonumber 
\end{equation}
Last equalities of Eqs.~(\ref{ec}) imply that 
\begin{equation}
\{ q_k,p_k\}_{u_k,v_k} = J_k^{-1} = 2i j \hbar/(1+u_kv_k)^2. 
\end{equation}
Since the stability matrix ${\mathbf{M}}_{\rm can}$ in the new set of variables is defined by
\begin{equation}
\delta \mathbf r''= {\mathbf{M}}_{\rm can} 
\delta \mathbf r',
\end{equation}
one can use Eq.~(\ref{zzss}) to find that ${\mathbf{M}} = {\mathbf{T}}''\, {\mathbf{M}}_{\rm can} \left(\mathbf{T}'\right)^{-1}$. It follows that
\begin{equation}
\det{\mathbf M} = \frac{\det\mathbf T''}{\det\mathbf T'}
\det{\mathbf{M}}_{\rm can} = \frac{J_x''J_y''}{J_x'J_y'}=
\mathcal T,
\label{Mdet}
\end{equation}
where $\mathcal T$ is given by Eq.~(\ref{Tcal}).


%

\end{document}